\documentclass[12pt]{article}
\usepackage{epsfig}
\usepackage{graphicx}
\newcommand{\amu}{$a_{\mu}$}
\def\Title#1{\begin{center} {\Large {\bf #1} } \end{center}}
\usepackage{fancyhdr}
\begin{document}
\topskip 2cm

\Title{Latest on the muon g-2 from experiment}
\bigskip

\begin{raggedright}  

{\it \underline{Graziano Venanzoni}  \index{}
\footnote{Presented at Linear Collider 2011: Understanding QCD at Linear Colliders  in searching for old and new physics, 12-16 September 2011, ECT*, Trento, Italy}\\
Laboratori Nazionali di Frascati dell' INFN, Frascati, Italy\\
{\rm  graziano.venanzoni@lnf.infn.it}
}\\

\bigskip\bigskip
\end{raggedright}
\vskip 0.5  cm
\begin{raggedright}
{\bf Abstract} We review the latest
 experimental achievements on the hadronic cross section measurements
at low energy which are of fundamental importance for a precise evaluation of the hadronic contribution to the $g$$-$$2$ of the muon.
We also discuss the new proposed muon $g$$-$$2$ experiments, with particular emphasis on E989 at Fermilab which plans 
to improve the experimental uncertainty by a factor of 4 with respect to the previous E821 experiment at BNL.
\end{raggedright}

\section{The muon anomaly as a precision test of the Standard Model\label{I}}

\begin{figure}[htb]
\begin{center}
\includegraphics[width=10cm,angle=0]{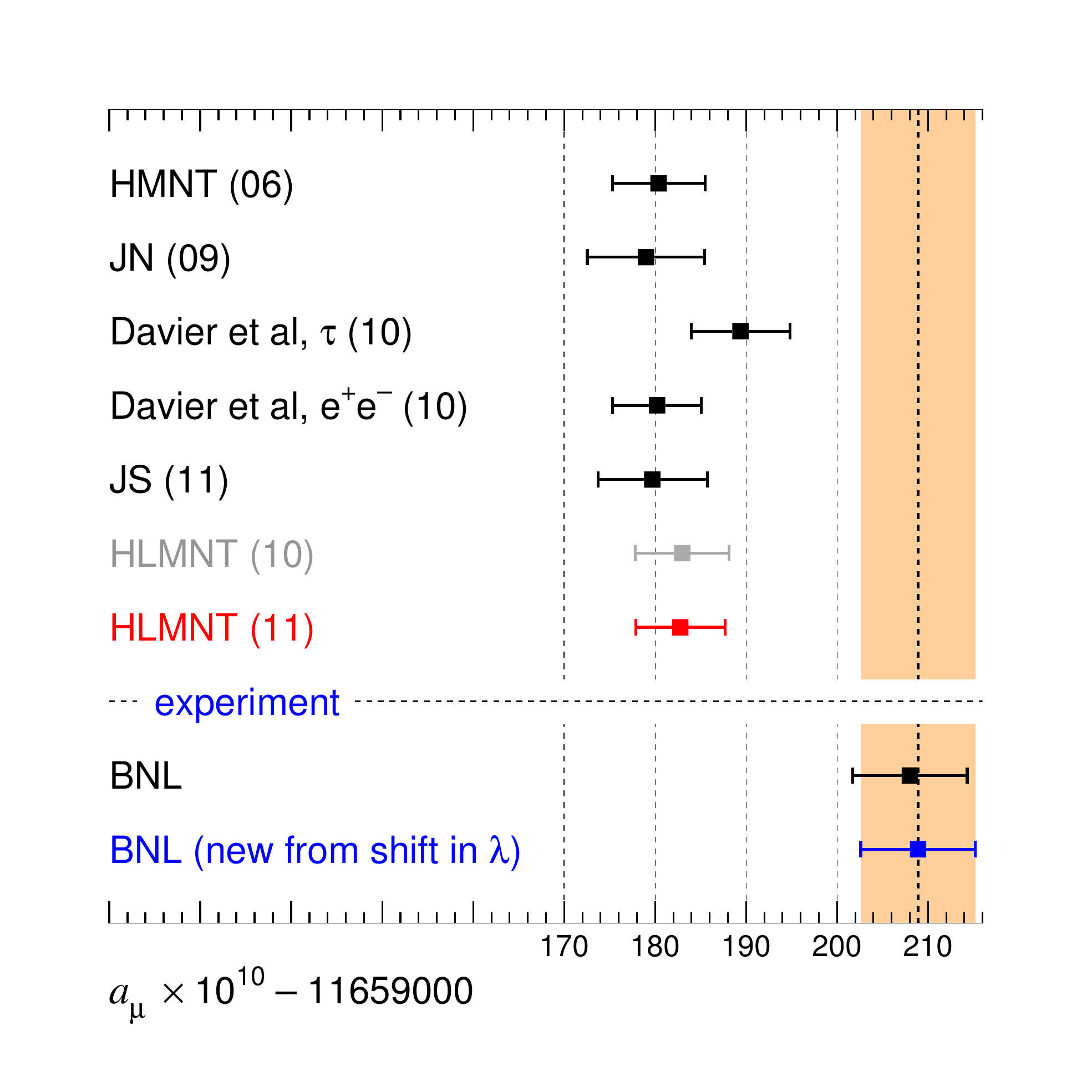}
\vspace{-1.cm}
\caption{Standard Model predictions of $a_\mu$ by several groups
  compared to the measurement from BNL (from Ref.~\cite{Hagiwara:2011af}).}
\label{fig1}
\end{center}
\end{figure}

The muon anomaly $a_{\mu}=(g-2)/2$ is a low-energy observable, which can be both measured and computed to high precision~\cite{Jegerlehner:2008zza}. Therefore it provides an important test of the Standard Model (SM) and allows a  sensitive search for new physics~\cite{Stockinger:1900zz}.
Since the first precision measurement of $a_{\mu}$ from the E821 experiment at BNL in 2001~\cite{Brown:2001mga}, 
there has been a  discrepancy between its
experimental value and the SM prediction.
This discrepancy
has been slowly growing due to recent impressive theory and experiment achievements. 
Figure~\ref{fig1} (from Ref.~\cite{Hagiwara:2011af}) shows an up-to-date  comparison of the
SM predictions by different groups and the BNL measurement for $a_{\mu}$.  
Evaluations of different groups are in very good agreement, showing a persisting $3\,\sigma$ discrepancy  (as, for example, $26.1\pm 8.0\times 10^{-10}$\cite{Hagiwara:2011af}).
It should be noted that both theoretical and experimental uncertainties have been reduced by more
than a factor of two in the last ten years\footnote{In 2001 this discrepancy was $(23.1\pm 16.9)\times 10^{-10}$~\cite{Prades:2001zv}.}.

\noindent
The accuracy of the theoretical prediction ($\delta a_{\mu}^{\rm{SM}}$,  
between 5 and 6 $\times 10^{-10}$)
  is limited by the strong interaction effects  which cannot be computed perturbatively at
low energies. Table~\ref{tab:g-2a} shows their contribution to the error for three recent 
estimates~\cite{Jegerlehner:2009ry,Davier:2010nc,Hagiwara:2011af}\footnote{Ref.~\cite{Jegerlehner:2009ry} uses a more conservative error analysis.}.
\begin{table}[h]
\begin{center}
 \renewcommand{\arraystretch}{1.4}
 \setlength{\tabcolsep}{1.6mm}
{\footnotesize
\begin{tabular}{|c|c c c|c|}
\hline
 Error & \cite{Jegerlehner:2009ry} & \cite{Davier:2010nc} & \cite{Hagiwara:2011af} & prospect \\
\hline
 $\delta a_{\mu}^{\rm
SM}$& 6.5 & 4.9 & 4.9 & 3.5 \\
\hline
 $\delta a_\mu^{\rm HLO}$ & 5.3 & 4.2 & 4.3 & 2.6 \\
$\delta a_\mu^{\rm HLbL}$ & 3.9 & 2.6 & 2.6 & 2.5 \\
\hline
$\delta (a_\mu^{\rm SM} - a_\mu^{\rm EXP})$ & 8.8 & 8.0 & 8.0 & 4.0 \\
\hline   
\end{tabular}
}
\caption{\label{tab:g-2a} Estimated uncertainties $\delta a_{\mu}$ in units
of $10^{-10}$ according to Refs.~\cite{Jegerlehner:2009ry,Davier:2010nc,Hagiwara:2011af} and
(last column) prospects in case of improved precision in the $e^+e^-$
hadronic cross section measurement (the prospect on $\delta a_\mu^{\rm HLbL}$ is an {\it educated guess}).
 Last row: Uncertainty on $\Delta a_{\mu}$ assuming the present experimental error of 6.3 from BNL-E821~\cite{Bennett:2006fi} (first two
columns) and of 1.6 (last column) as planned by the future 
($g$$-$$2$) experiments~\cite{Carey:2009zz,Imazato:2004fy}.}
\end{center}
\end{table}
 The leading-order hadronic vacuum polarization 
contribution,  $a_{\mu}^{\rm{HLO}}$, gives the main uncertainty (between 4 and 5 $\times 10^{-10}$).
It can be related by dispersion integral to the measured hadronic cross sections, and 
it is known with a fractional accuracy of 0.7\%, i.e. to about 0.4 ppm.
The O($\alpha^3$) hadronic light-by-light  contribution, $a_{\mu}^{\rm{HLbL}}$,
 is the second dominant error in the theoretical evaluation. 
It cannot at present be determined from data, and relies on specific models. 
Although its value is almost one order of magnitude smaller than  $a_{\mu}^{\rm{HLO}}$, it is much worse known 
(with a fractional error of the order of 30\%)
and therefore it still gives
a significant contribution to $\delta a_{\mu}^{\rm{SM}}$ (between 2.5 and 4 $\times 10^{-10}$).
From the experimental side, the 
error achieved by the BNL E821 experiment  is $\delta a_{\mu}^{\rm{EXP}}= 6.3 \times 10^{-10}$ (0.54 ppm)~\cite{Bennett:2006fi}. 
This impressive result is still limited by the statistical error,
and experiments to measure the muon $g$$-$$2$ with a fourfold improvement in accuracy 
have been approved at Fermilab~\cite{Carey:2009zz} and J-PARC~\cite{Imazato:2004fy}.

\section{Recent progress on the hadronic contribution to \amu}
Differently from the QED and Electroweak contributions to \amu, which can be calculated
using perturbation theory, and therefore are well under
control, the hadronic ones (LO VP and HLbL) 
cannot be computed reliably using perturbative QCD.  
The lowest order hadronic contribution  $a_{\mu}^{\rm{HLO}}$ can be computed from hadronic $e^+ e^-$ annihilation data 
via a dispersion relation, and therefore its uncertainty strongly depends on the accuracy of the experimental 
data. For the hadronic Light-by-Light contribution  $a_{\mu}^{\rm{HLbL}}$ there is no direct connection with data and therefore only model-dependent estimates exist.
As  the  hadronic sector dominates the uncertainty on the theoretical prediction  $a_{\mu}^{\rm{SM}}$, 
considerable effort has been put  on it by experimental and theoretical groups, reaching the following main results:
\begin{itemize}
\item A precise determination of the hadronic cross sections at the $e^+e^-$ colliders (VEPP-2M, DA$\mathrm{\Phi}$NE, BEPC, PEP-II and KEKB) which allowed a determination of  $a_{\mu}^{\rm{HLO}}$ 
with a fractional error below 1\%.
These efforts led to the development of dedicated high precision theoretical tools, like the inclusion of high-order Radiative
 Corrections (RC) and the 
non-perturbative hadronic contribution to the running of $\alpha$ 
(i.e. the vacuum polarisation, VP)
in Monte Carlo (MC) programs used for the analysis of the data~\cite{Actis:2010gg};


\item Use of \emph{Initial State Radiation} (ISR)~\cite{Chen:1974wv,Binner:1999bt,Benayoun:1999hm} which opened a new way to precisely obtain
 the electron-positron annihilation  cross sections into hadrons at particle factories 
operating at fixed beam-energies~\cite{Kluge:2008fb,Druzhinin:2011qd};

\item A dedicate effort on the evaluation of the Hadronic Light-by-Light contribution, where 
two different groups~\cite{Prades:2009tw,Jegerlehner:2009ry} 
found agreement on the size of the contribution (with slightly different errors), and therefore strengthening our confidence in the reliability of these estimates;
\item An impressive progress on QCD calculation on the lattice, where 
an accuracy better than $3\%$ was reached on the two-flavor QCD correction to  $a_{\mu}^{\rm{HLO}}$~\cite{Feng:2011zk};

\item Better agreement between the $e^+e^-$ and the $\tau$ based evaluation of  $a_{\mu}^{\rm{HLO}}$, thanks to improved isospin corrections~\cite{Davier:2010nc}. 
These two sets of data are eventually in agreement (with $\tau$ data moving towards 
$e^+e^-$  data) after including vector
meson and $\rho - \gamma$  mixing \cite{Jegerlehner:2011ti,Benayoun:2011mm}.
\end{itemize}
\vspace{0.5cm}


 %
%


\section {$\sigma_{had}$ measurements at low energy}
\label{sec:rstatus}
\begin{figure}[h]
\begin{center}
\resizebox{0.6\columnwidth}{!}{
\includegraphics[width=8cm,height=7cm]{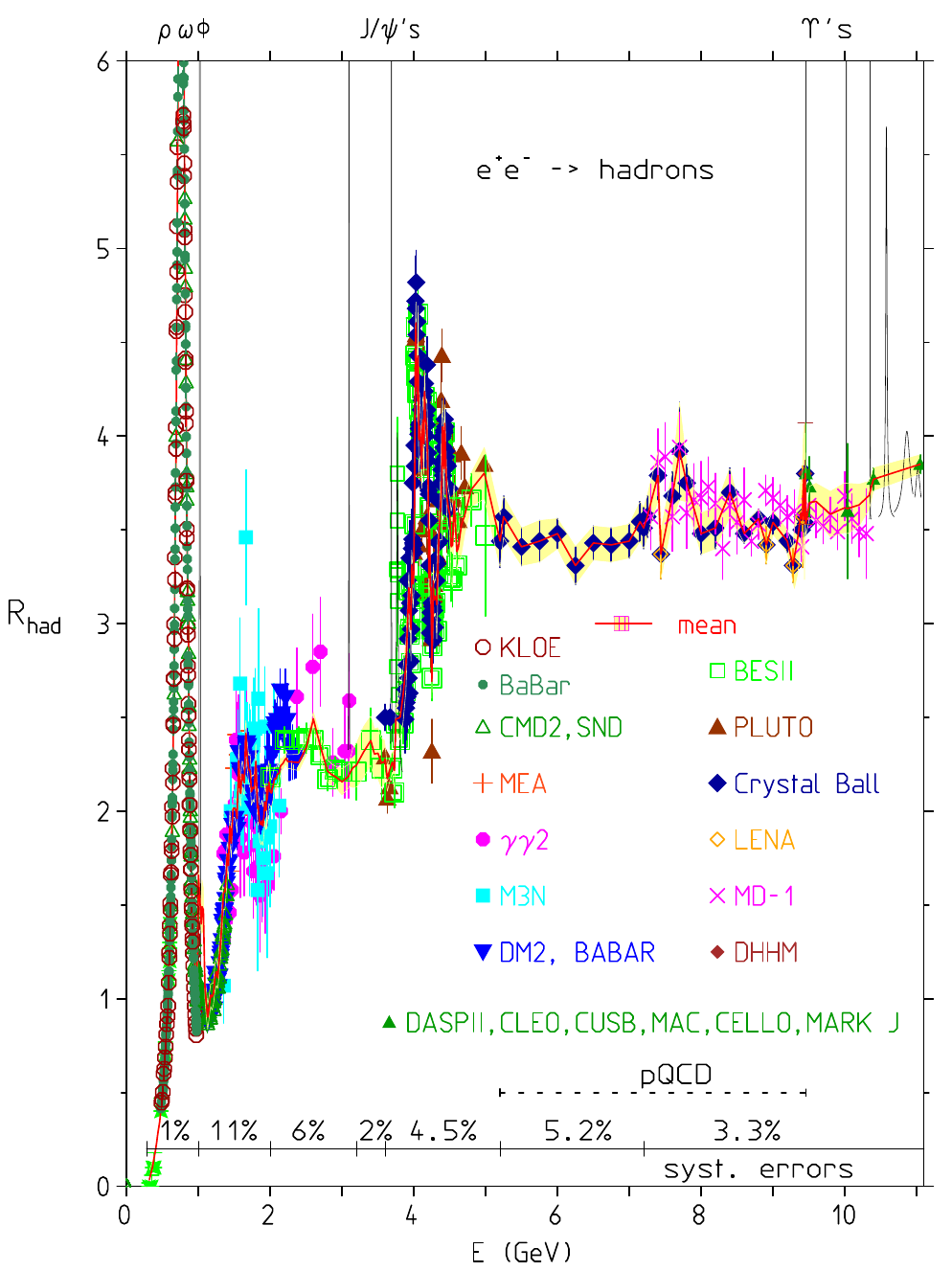}}
\vspace{-.5cm}
\caption{An updated compilation of $R$ measurements.
 In the bottom line the overall uncertainties of the different
regions are reported ({\it courtesy of Fred Jegerlehner}).} 
\label{rfj}
\end{center}
\end{figure}

In the last few years, big efforts on $e^+e^-$ data
in the energy range below a few GeV led to a substantial reduction 
in the hadronic uncertainty on  $a_{\mu}^{\rm HLO}$.
Figure \ref{rfj} shows an up-to-date compilation of these data.
The main improvements have been achieved
 in the region below 5 GeV: between 2 and 5 GeV 
(where the data are now closer to the prediction of 
pQCD), the BESII collaboration reduced the error 
to $\sim$7\%~\cite{Bai:2001ct}
(before it was $\sim$15\%); between 1 and 4.5 GeV 
BaBar measured various final states with more than two hadrons with a
 systematic accuracy between 3\% and 15\%, 
as shown in Tab. \ref{babar_r};
below 1 GeV, the CMD-2~\cite{Akhmetshin:2006bx,Akhmetshin:2006wh,Akhmetshin:2003zn} and 
SND~\cite{Achasov:2006vp} collaborations at Novosibirsk, 
KLOE~\cite{Ambrosino:2010bv,:2008en,Aloisio:2004bu} at Frascati and BaBar~\cite{:2009fg} at Stanford  measured
the pion form factor in the energy range around the $\rho$ peak with a
systematic error of $0.8\%$, $1.3\%$, $0.9\%$, and $0.5\%$, respectively.
\begin{table}
\vspace{2mm}
\label{babar_r}
\begin{center}
\begin{tabular}{|c|c|}
\hline
Process & Systematic accuracy \\
\hline
\hline
$\pi^+\pi^-\pi^0$ &(6-8)\%  \\
$2\pi^+2\pi^-$ & (3-8)\%  \\  
$2\pi 2\pi^0$  & (8-14)\% \\
$2(\pi^+\pi^-)\pi^0,2(\pi^+\pi^-)\eta$& (7-10)\%  \\
$3\pi^+3\pi^- , 2\pi^+2\pi^-2\pi^0$ & (6-11)\%   \\
$KK\pi,KK\eta$  & (5-6)\% \\
$K^+K^-\pi\pi$  & (8-11)\% \\
$K^+K^-\pi^+\pi^-\pi^0$, $K^+K^-\pi^+\pi^-\eta$ & (5-10)\%  \\
$2(K^+K^-)$ & (9-13)\% \\
\hline
\end{tabular}
\caption{Systematic accuracy on more than two hadrons 
processes studied by 
BaBar 
in the energy range 1$< \sqrt{s} <$ 4.5 GeV using ISR. 
}
\end{center}
\end{table}

\begin{figure}[ht]
\vspace*{-7mm}
\centering
\includegraphics[height=5cm]{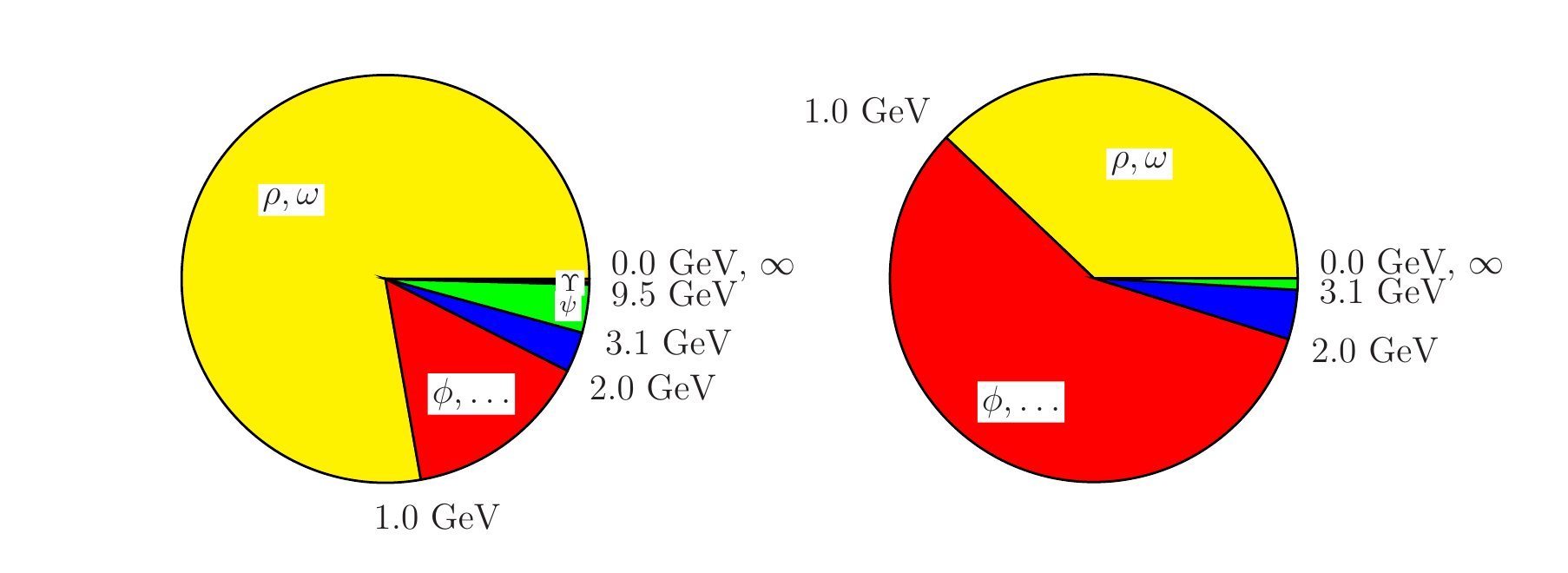}
\vspace{-.5cm}
\caption{The distribution of contributions (left) and errors (right)
in \% for $a_{\mu}^{\mbox{$\scriptscriptstyle{\rm HLO}$}}$ 
from different energy regions. The error of a
contribution $i$ shown is
$\delta^2_{i\:{\rm tot}}/\sum_i \delta^2_{i\:{\rm tot}}$ in \%. The
total error combines statistical and systematic errors in quadrature 
(from Ref.~\cite{Jegerlehner:2009ry}).}
\label{fig:gmusta}
\end{figure}

The CMD-2 and SND collaborations at Novosibirsk and BESII in Beijing  
were performing the hadronic cross section measurements in a 
traditional way, i.e., by varying the $e^+e^-$ beam energies.
KLOE, BaBar, and more recently Belle 
used ISR (also called {\it radiative return}) 
as reviewed in Refs. \cite{Actis:2010gg,Kluge:2008fb,Druzhinin:2011qd}.
Figure \ref{rfj} shows that, despite the recent progress, 
the region between 1 and 2 GeV is still poorly known, with a fractional 
accuracy of $\sim$6\%. Since about 50\% of the error squared, 
$\delta^2 a_{\mu}^{\rm HLO}$ comes from this region (see Fig.~\ref{fig:gmusta}),   
it is 
evident how desirable an improvement on the hadronic cross section
 of this region is.


\begin{figure}[ht]
\centering
\includegraphics[height=8cm]{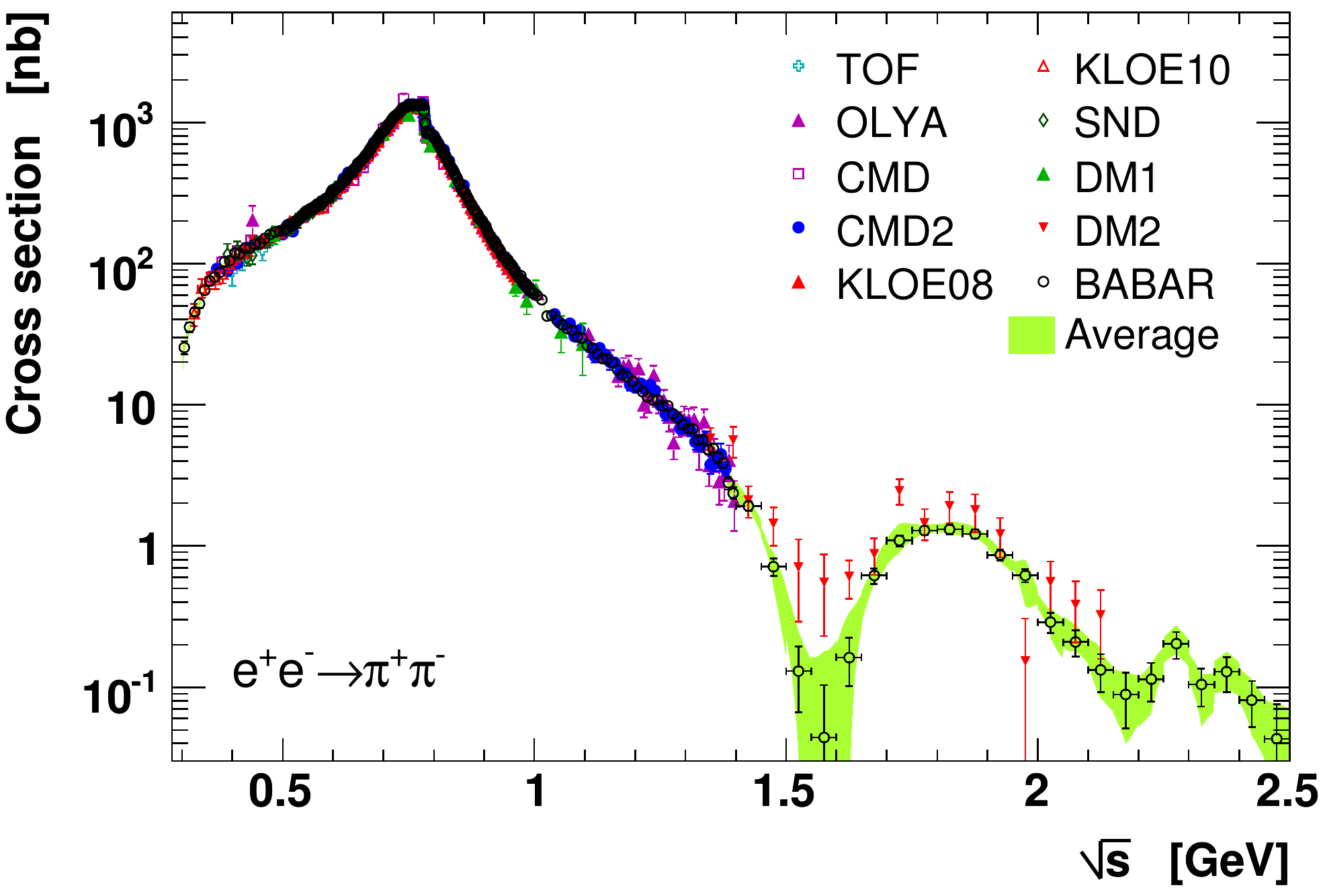}
\vspace{-.5cm}
\caption{ Cross section of $e^+e^-\to\pi^+\pi^-$  as measured  by different experiments (from Ref.~\cite{Davier:2010nc}).}
\label{fig:davier}
\end{figure}

\subsection{Measurement of  $\sigma_{\pi \pi}$ below 1 GeV}
\label{sec:lowhadcs}
The region below 1 GeV is dominated by the two-pion channel which 
accounts for 70\% of the contribution to $a_{\mu}^{\rm HLO}$, and for
40\% to the total squared error of $a_{\mu}$ (see Fig.~\ref{fig:gmusta}).
\noindent
Therefore due to its particular importance, 
it has been studied by different experiments as shown in Fig.~\ref{fig:davier}.
CMD-2
and SND
have performed an energy
scan at the $e^+e^-$ collider VEPP-2M ($\sqrt{s}\in$ [0.4--1.4]
GeV) with $\sim$10$^6$ and $\sim$4.5$\times 10^6$ events respectively, and
systematic fractional errors from 0.6\% to 4\% in the cross sections,
depending on $\sqrt{s}$. 
The pion form factor has also been 
measured by KLOE and more recently by 
BaBar, both using ISR.
KLOE collected more than 3.1 million events, corresponding
 to an integrated luminosity of 240 pb$^{-1}$, leading to a relative error
of 0.9\% in the energy region [0.6--0.97] GeV dominated by systematics.
BaBar has performed a $\pi^+\pi^-(\gamma)$ cross section 
measurement based on half a million selected events. 
The pion form factor is obtained by the ratio 
$\pi^+\pi^-(\gamma)$ to $\mu^+\mu^-(\gamma)$ which allows 
a systematic error of 0.5\% in the $\rho$ region 
increasing to 1\% outside.
The threshold region [2$m_\pi - 0.5$ GeV] provides 13\% of the total
$\pi^+\pi^-$ contribution to the muon anomaly: 
$a_\mu^{\rm HLO}$ [2$m_\pi - 0.5$ GeV] = $(58.0\pm2.1) 
\times$ 10$^{-10}$. To overcome the lack of
precision data at threshold energies, 
the pion form factor 
is extracted from a
parameterization based on ChPT, constrained by
spacelike data~\cite{Amendolia:1986wj}.
The most effective way to measure the cross section near the 
threshold in the timelike region is
provided by ISR events, where the emission of an energetic photon allows
to study the two pions at rest.  
BaBar has achieved an error between 0.8 and 1.4\% in this region, 
while KLOE has achieved a larger error (up to 7\%) dominated by  the point-like model uncertainty for FSR.

\noindent
There is a fair agreement between the four experiments in the region below 1 GeV, 
with a discrepancy of about 2-3\% between KLOE (lowest cross section) and BaBar  (highest cross section) at the $\rho$ peak,
 and CMD2 and SND somehow in the middle. 
Although small, this difference is larger than the claimed systematic error 
and can be a limitation for further improvements of  $a_{\mu}^{\rm SM}$. 
As BaBar and KLOE (published) data use a different normalization (to muon pair and to Bhabha events,  respectively) it may be that part of this difference can come from the normalization procedure itself.
In order to check this possibility, the KLOE experiment has 
recently presented a new {\it preliminary}  measurement of the pion form factor derived from the bin-by-bin $\pi^+\pi^-\gamma/\mu^+\mu^-\gamma$ ratio~\cite{eps} as done in BaBar.
\begin{figure}[h]
\begin{center}
\resizebox{0.8\columnwidth}{!}{
\includegraphics[width=8cm]{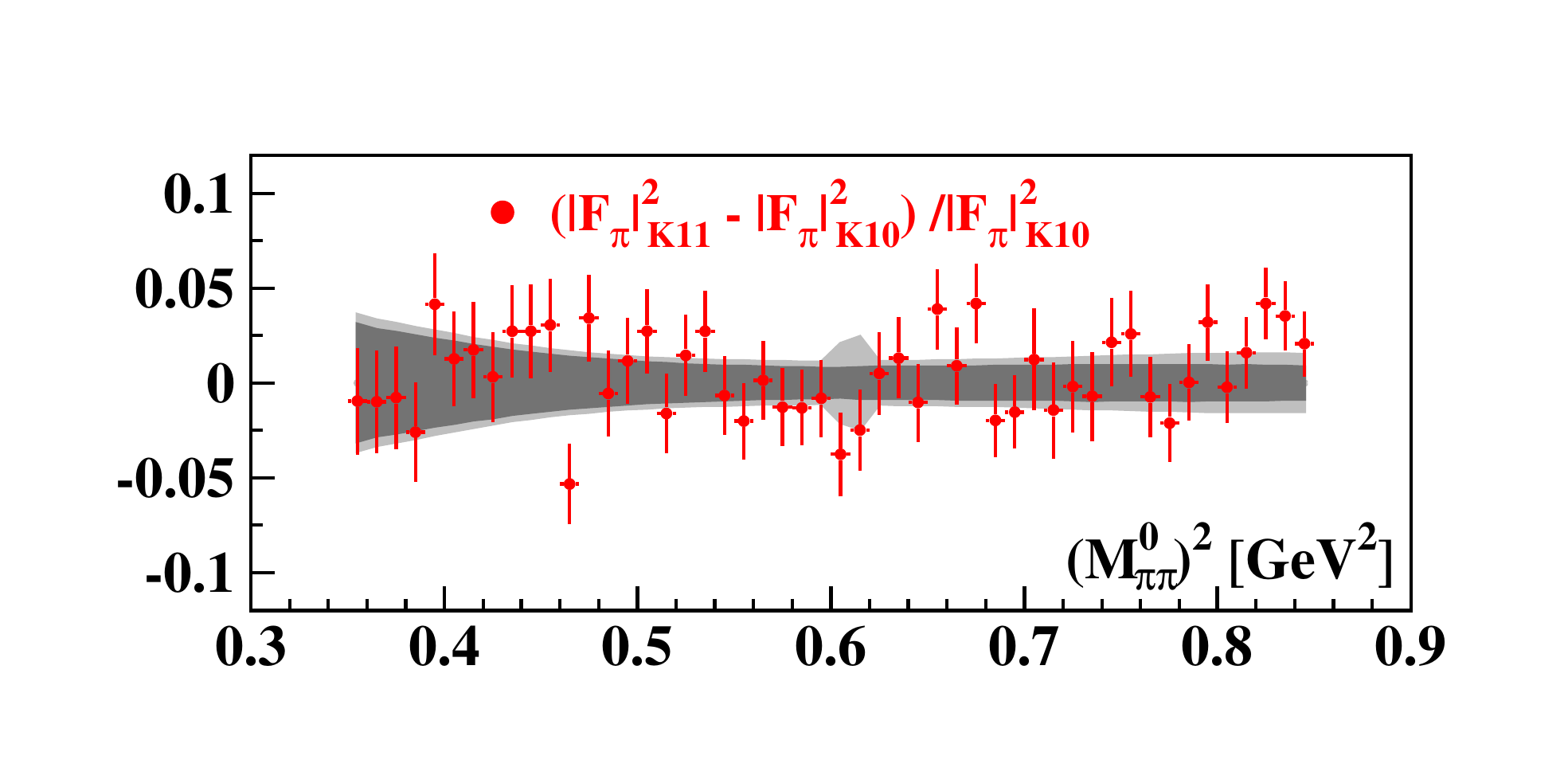}}
\vspace{-1.cm}
\caption{Fractional difference between the published KLOE measurement normalized to Bhabha events~\cite{Ambrosino:2010bv} and the new {\it preliminary} one derived from the bin-by-bin $\pi^+\pi^-\gamma/\mu^+\mu^-\gamma$ ratio.}
\label{kloe2}
\end{center}
\end{figure}
As can be shown in Fig.~\ref{kloe2}, good agreement is found between the two spectra, which excludes possible problems in the normalization procedure used in KLOE.
%

\subsection{Measurement of  $\sigma_{had}$ above 1 GeV}
\label{sec:highhadcs}
The region [1--2.5] GeV, with an uncertainty on $\sigma_{had}$ between 6 and 10\%, is the most poorly known, and contributes about 
55\% of the uncertainty on $a_{\mu}^{\mbox{$\scriptscriptstyle{\rm HLO}$}}$ (see Fig.~\ref{fig:gmusta}).
In this region BaBar using ISR has published results  on 
$e^+ e^-$ into three, four, five and six hadrons, with a
 general improvement  with respect to the much less precise measurements
from M3N, DM1 and DM2. For several channels, BaBar measured lower cross sections with respect to older experiments, resulting in a reduced contribution from this energy region to $a_{\mu}^{\mbox{$\scriptscriptstyle{\rm HLO}$}}$.
Recently CMD-3 and SND experiments at the
upgraded VEPP-2000 collider in Novosibirsk have presented new measurements on multihadron channels~\cite{solodov}. With about 20 pb$^{-1}$ of collected data, they have achieved a 
statistical error comparable to ISR data from B-factories.
VEPP-2000 plans to collect an integrated luminosity of $~1$ fb$^{-1}$, which would allow a significant improvement for many channels in the region below 2 GeV.\\
With a specific luminosity of $10^{32}$cm$^{-2}$s$^{-1}$,
   DA$\mathrm{\Phi}$NE upgraded in energy, could perform 
a scan in the region from 1 to 2.5 GeV,  
collecting 
an integrated luminosity of 20 pb$^{-1}$ per point corresponding 
to few days of data taking for each energy bin~\cite{Babusci:2010ym}. 
By assuming an energy step of 25 MeV, the whole
region would be scanned in one year of data taking.
The statistical yield 
 would be one order of magnitude higher 
than what would have been achieved with 1 ab$^{-1}$ at BaBar, and better than 
what is to be expected at BESIII with 10 fb$^{-1}$ at 3 GeV.

\noindent
Finally, prospects of reaching an integrated luminosity by a factor
of 30-100 exceeding that of the present machines appear at Super B-Factories.
Such machines
will improve accuracy for many processes whose studies
are now statistically limited.


\section{Measuring \amu}

The muon anomaly \amu\ has been measured with better and better accuracy during the last 50 years.  The E821 experiment at Brookhaven has reached an impressive 14-fold improvement in precision with respect to the pioneering measurements performed
at CERN. Two new experiments  with a goal of fourfold improvement in accuracy are underway: the approved E989  at Fermilab~\cite{Carey:2009zz}, and the J-PARC proposal~\cite{Imazato:2004fy} that has recently received stage-one approval.
E989 is based on the well known magic-momentum concept and uses the BNL storage ring as a key element. The proposal at J-PARC uses a new approach with ultra-slow muons at off-magic momentum.
We will now discuss how the measurement of \amu\ is done, describing 
the E821 experiment, and its upgrade E989.

\noindent
The measurement of \amu\ uses the spin precession
resulting from the torque experienced by
the magnetic moment when placed in a magnetic
field. An ensemble of polarized muons is introduced
into a magnetic field, where they are stored
for the measurement period. 
The rate at which the spin rotates
relative to the momentum vector is given by the
difference in frequency between the spin precession
and cyclotron frequencies. 
Because electric quadrupoles are used to provide vertical focusing in the storage ring, their
electric field is seen in the muon rest frame as a moving magnetic field that can affect the
spin precession frequency.
In the presence of both $\vec{E}$ 
and $\vec{B}$ fields, and in the case that $\vec{\beta}$
is perpendicular to both, the anomalous precession frequency ({\it i.e.} the frequency at which 
the muon’s spin advances relative to its
momentum)
is 
\begin{eqnarray}
\nonumber
\vec{\omega_a} & =& \omega_S - \omega_C \\
 & = & -\frac{q}{m}\Big [a_{\mu}\vec{B} - \Big (a_{\mu} - \frac{1}{\gamma^2-1}\Big )\frac{\vec{\beta}\times \vec{E}}{c}\Big ]\label{eq1}
\end{eqnarray}

\noindent
The experimentally measured
numbers are the muon spin frequency $\omega_a$ and the
magnetic field, which is measured with proton
NMR, calibrated to the Larmor precession 
frequency, $\omega_p$, of a free proton. The anomaly is related
to these two frequencies by
\begin{eqnarray}
a_{\mu} & = & \frac{\tilde{\omega_a}/\omega_p}{\lambda-\tilde{\omega_a}/\omega_p} = 
\frac{R}{\lambda - R},
\end{eqnarray}
where
$\lambda = \mu_\mu/\mu_p = 3.183 345 137(85)$ (determined experimentally
from the hyperfine structure of muonium), and  
$ R = \tilde{\omega_a}/\omega_p$ .
The tilde over $\omega_a$ means that it has been corrected
for the electric-field and pitch ($\vec{\beta}\cdot\vec{B} \neq  0$)
corrections [3]. 
The magnetic field in Eq.~\ref{eq1}
is an average that can be expressed as an integral
of the product of the muon distribution times
the magnetic field distribution over the storage
region. Since the moments of the muon distribution
couple to the respective multipoles of the
magnetic field, either one needs an exceedingly
uniform magnetic field, or exceptionally good information
on the muon orbits in the storage ring,
to determine the $<B_\mu>$ distribution to sub-ppm precision.
This was possible in E821 where the
uncertainty on the magnetic field averaged over
the muon distribution was 30 ppb (parts per
billion). 
The coefficient of the $\vec{\beta}\times\vec{E}$ 
term in Eq.~\ref{eq1} vanishes at the ``magic" momentum of 3.094 GeV/c;
where $\gamma = 29.3$. Thus \amu\ can be determined by a precision measurement of $\omega_a$ and B. 
At this magic momentum, the electric field  is used only for muon storage and the magnetic 
field alone determines the precession frequency. The finite spread in beam momentum
and vertical betatron oscillations introduce small (sub ppm) corrections to the precession
frequency. These are the only corrections made to the measurement.

\noindent The experiment consists of repeated fills of the
storage ring, each time introducing an ensemble
of muons into a magnetic storage ring, and then
measuring the two frequencies $\omega_a$ and $\omega_p$. The
muon lifetime at the magic momentum is 64.4 $\mu$s, and the
data collection period is typically  700 $\mu$s. The $g$$-$$2$ precession period 
is 4.37 $\mu$s, and the cyclotron period $\omega_C$  149 ns.

\begin{figure}[htb]
\begin{center}
\includegraphics[width=8cm,angle=0]{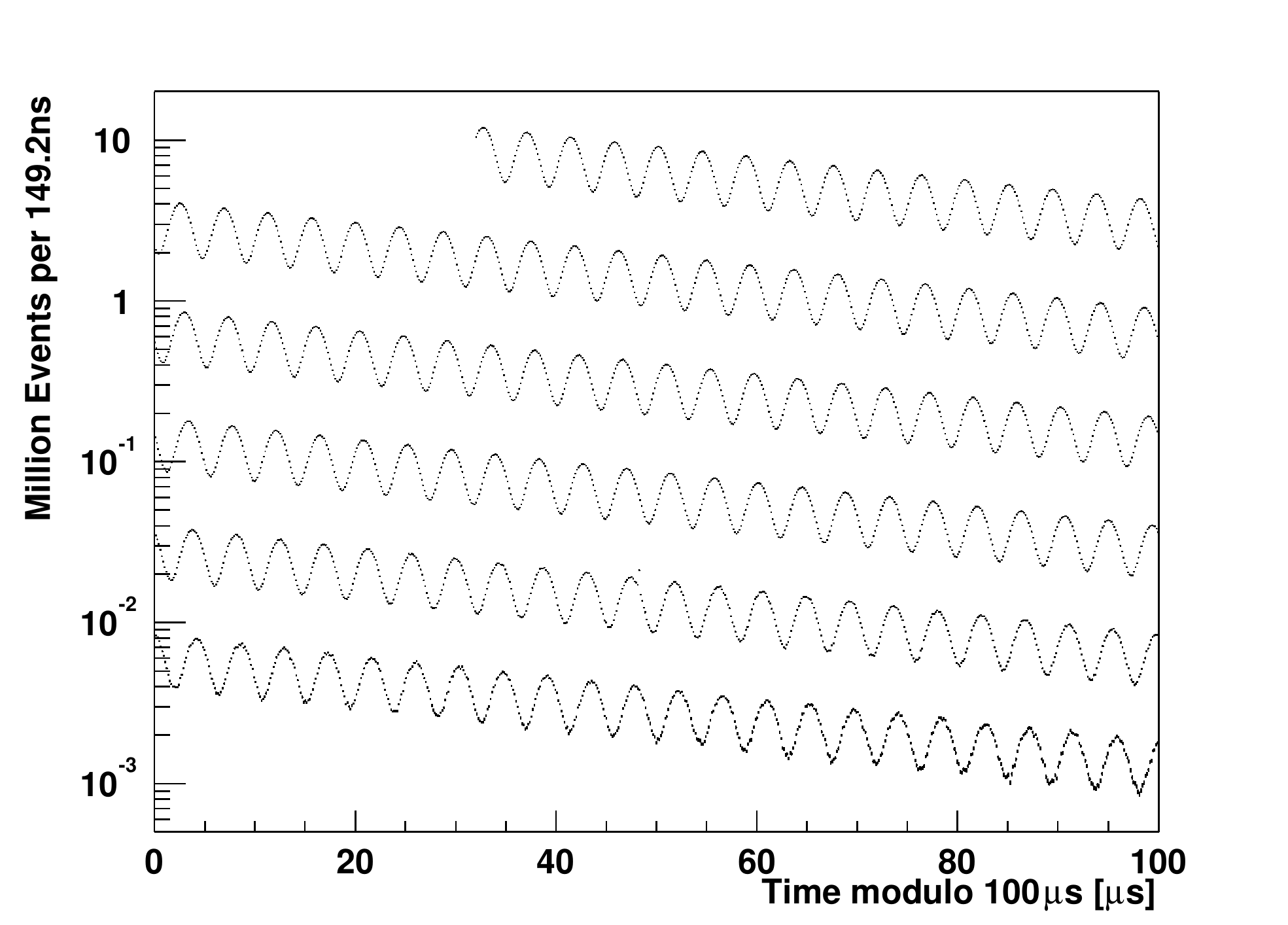}
\caption{Distribution of electron counts versus time for the 3.6 billion muon decays. The data are
wrapped around modulo 100 $\mu$s~\cite{Bennett:2006fi}.}
\label{fig2}
\end{center}
\end{figure}

\noindent Because of parity violation in the weak decay of the muon, a correlation exists between
the muon spin and the direction of the high-energy decay electrons.
Thus as the spin rotates relative to the momentum,
the number of high-energy decay electrons
is modulated by the frequency $\omega_a$, as shown
in Fig.~\ref{fig2}.
The E821 storage ring was constructed as a
“super-ferric” magnet, meaning that the iron
determined the shape of the magnetic field. Thus
the magnetic field needed to be well below saturation and was
chosen to be 1.45 T. The resulting ring had a 
central orbit radius of 7.112 m, and 24 detector stations
were placed symmetrically around the inner
radius of the storage ring. 
The detectors were made of Pb/SciFi electromagnetic calorimeters 
which measured the decay electron energy and time of arrival.
The detector geometry
and number were optimized to detect the high energy
decay electrons, which carry the largest
asymmetry, and thus information on the muon
spin direction at the time of decay. In this design,
many of the lower-energy electrons miss the
detectors, reducing background and pileup.


\section{The Fermilab proposal: E989}
The E989 collaboration at Fermilab plans to measure \amu\ with
 an uncertainty of $1.6\times 10^{-10}$ (0.14 ppm), 
corresponding to a 0.10 ppm
statistical error and roughly equal 0.07 ppm systematic uncertainties on $\omega_a$ and $\omega_p$.


\noindent The proposal efficiently uses the unique properties of the Fermilab beam
complex to produce the necessary flux
of muons, which will be injected and stored in the (relocated) muon storage ring. 
To achieve a statistical uncertainty of 0.1 ppm, the total data set must contain more than
$1.8\times 10^{11}$ detected positrons with energy greater than 1.8 GeV, and arrival time greater
than 30 $\mu$s after injection into the storage ring. The plan uses 6 out of 20 of the 8-GeV
Booster proton batches, 
each subdivided into four bunches of $10^{12}$ p/bunch.
 The proton bunches fill the muon storage ring at a repetition rate
of 15 Hz, to be compared to the 4.4 Hz at BNL.
 The proton bunch hits a target,
 producing a 3.1 GeV/c pion beam that is directed along a greater than 1 km decay line.
The resulting pure muon beam is injected into the storage ring.
The muons will enter the ring through a new superconducting inflector magnet, which will replace the 
existing one, which is wound in such a manner that the coils intercept the beam on both ends
of the magnet. The new inflector will result 
in a higher muon storage efficiency. Once entering the ring, an optimized
pulse-forming network will energize the storage ring kicker to place the beam on a stable
orbit. 
The pion flash (caused by pions entering the ring at injection) will be decreased by a factor of 20
from the BNL level, and the muon flux will be significantly
increased because of the ability to take
zero-degree muons. The stored muon-per-proton
ratio will be increased by a factor of 5 to 10 over
BNL. 

\noindent The E821 muon storage will be relocated
to Fermilab, in a new
building with a stable floor and good temperature
control, neither of which were available at
Brookhaven.


\noindent The new experiment will require upgrades of detectors, electronics and data acquisition
equipment to handle the much higher data volumes and slightly higher instantaneous rates.
High-density segmented tungsten/scintillating-fibers~\cite{McNabb:2009dz} and crystals are considered as possible choice for the calorimeter.
In-vacuum straw drift tubes have been developed to determine the stored muon distribution from decay positron tracks and to provide data for a greatly improved muon electric
dipole moment measurement, which can be obtained in parallel~\cite{Bennett:2008dy}.
 A modern data acquisition system will be used
to read out waveform digitizer data and store it so that both the traditional event mode
and a new integrating mode of data analysis can be used in parallel. 
The systematic uncertainty on the precession frequency is expected to improve by a factor 3 
thanks to the reduced pion contamination,
the segmented detectors, and an improved storage ring kick of the muons onto orbit.
The storage ring magnetic field will be shimmed to an even more impressive uniformity,
and improvements in the field-measuring system will be implemented. 
The systematic error on the magnetic field
is halved by better shimming, relocations of critical NMR probes, and other incremental
changes.

\noindent 
In less
than two years of running, the statistical goal of $4\times 10^{20}$ protons on target can be achieved for positive muons. A follow-up run using negative muons is possible, depending on future scientific motivation. Two additional physics results will be obtained from the same data: a new limit on the muon's
electric dipole moment (up to 100 times better); and, a more stringent limit on possible
CPT or Lorentz violation in muon spin precession. A technically driven schedule permits
data taking to begin in 2016.

\section{Prospects on \amu}
With the new experiments planned at Fermilab and J-PARC the uncertainty of the difference $\Delta a_{\mu}$ between the experimental and the
theoretical value of 
$a_{\mu}$ will be dominated by the uncertainty of the hadronic
cross sections at low energies, unless new experimental efforts
 at low energy are undertaken.
 The last column of Table~\ref{tab:g-2a} shows
a future scenario based on
realistic improvements in the $e^+e^-\to hadrons$  cross sections measurements. Such
improvements could be obtained by reducing the uncertainties of the hadronic
cross sections from 0.7\% to 0.4\% in the region below 1 GeV and from 6\%
to 2\% in the region between 1 and 2 GeV as shown in Table~\ref{tab:g-2b}.

\begin{figure}[ht]
\begin{center}
\includegraphics[height=9cm,width=9cm]{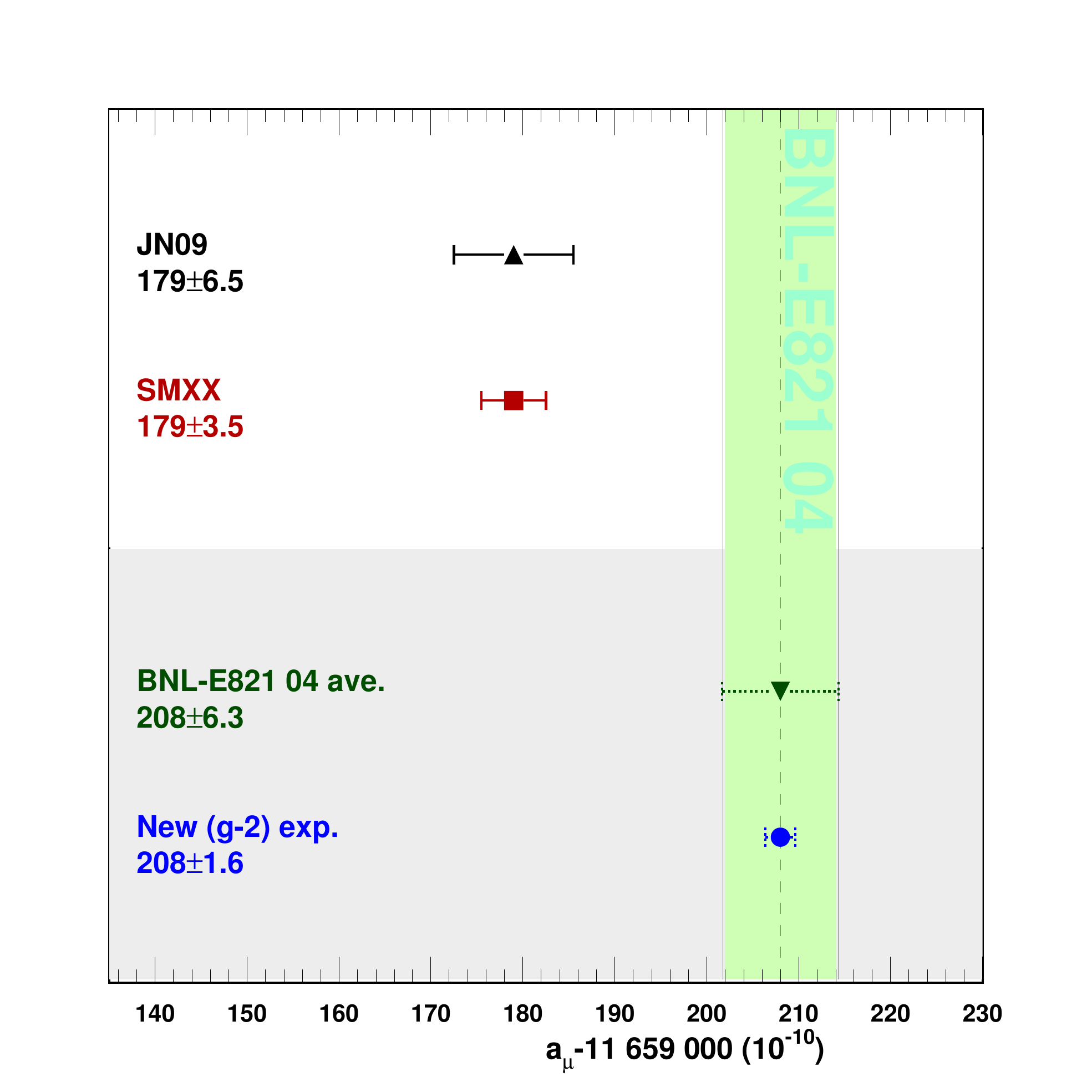}
\vspace{-.5cm}
\caption{Comparison between $a_{\mu}^{\rm SM}$ and $a_{\mu}^{\rm EXP}$. 
``JN09'' is the current evaluation of $a_{\mu}^{\rm SM}$ using Ref.~\cite{Jegerlehner:2009ry}; ``SMXX'' is the same central value with a reduced error as
 expected by the improvement on the hadronic cross section measurement (see text); ``BNL-E821 04 ave.'' is the current experimental value
of $a_{\mu}$;  ``New ($g$$-$$2$) exp.'' is the same central value with a fourfold improved accuracy as planned by the future ($g$$-$$2$) experiments~\cite{Carey:2009zz,Imazato:2004fy}.}
\label{fig:g-2xx}
\end{center}
\end{figure}

\begin{table}[h]
\begin{center}
 \renewcommand{\arraystretch}{1.4}
 \setlength{\tabcolsep}{1.6mm}
{\footnotesize
\begin{tabular}{|c|c|c|c|c|}
\hline
 & $\delta (\sigma)/\sigma$ present &$\delta a_{\mu}^{\rm{HLO}}$ present
 & $\delta (\sigma)/\sigma$ prospect &$\delta a_{\mu}^{\rm{HLO}}$ prospect  \\
\hline
$\sqrt{s}<1$~GeV & 0.7\% & 3.3 & 0.4\% & 1.9 \\
$1<\sqrt{s}<2$~GeV & 6\% & 3.9 & 2\% & 1.3 \\
$\sqrt{s}>2$~GeV & & 1.2 & & 1.2 \\
\hline
total & & 5.3 & & 2.6 \\
\hline
   
\end{tabular}
}
\caption{\label{tab:g-2b} Overall uncertainty of the cross-section
measurement required to get the reduction of uncertainty on $a_{\mu}^{\rm{HLO}}$ in units
$10^{-10}$ for
three regions of $\sqrt{s}$ (from Ref.~\cite{Jegerlehner:2008zz}).}
\end{center}
\end{table}
\noindent In this scenario the overall uncertainty on $\Delta a_{\mu}$ could be
reduced by a factor 2. In case the central value would remain the same, the
statistical significance would become 7-8 standard deviations, as it can be seen in Fig.~\ref{fig:g-2xx}.

\noindent  The effort needed to reduce the uncertainties of the $e^+e^-\to hadrons$
cross-sections according to Table~\ref{tab:g-2b} is challenging but possible, and certainly
well motivated by the excellent opportunity the muon $g$$-$$2$ is
providing us to unveil (or constrain) ``new-physics'' effects. 
A long-term program of hadronic cross section measurements at low energies is clearly 
warranted and fortunately it has been already started at VEPP-2000.
In addition, recent theoretical activities  focused on lattice calculation have already 
reached a mature stage and have real 
prospects to match the future experimental precision.\\
With the expected reduction of the error on  $a_{\mu}^{\rm{HLO}}$, and the 
planned improved precision of the new $g$$-$$2$ experiments, the hadronic Light-by-Light contribution  
could become the main limitation for further progress on  $a_{\mu}^{\rm{SM}}$.
Although there isn't a direct connection with data,
 $\gamma\gamma$ measurements performed at $e^+ e^-$ colliders will help us 
to constrain form factors~\cite{Babusci:2011bg}.
Lattice calculation could help as well.

\section{Conclusion}
The measurements of the muon anomaly \amu\  have been a important benchmark
 for the development of QED and the Standard Model. 
In the recent years, following the impressive accuracy (0.54 ppm) reached by  
the E821 experiment at BNL,
a worldwide effort from different theoretical and experimental groups 
has significantly improved the SM prediction.
At present 
there appears to be a 3$\sigma$ difference between the experimental value and the SM prediction of $a_{\mu}$.
This discrepancy, which  would fit well with SUSY expectations, is a 
valuable constraint in restricting physics beyond the Standard Model, guiding  
the interpretation of  LHC results.
In order to clarify the nature of the observed discrepancy between theory and experiment,
and eventually firmly establish (or constrain) new physics effects,
new direct measurements of the
muon $g$$-$$2$ with a fourfold improvement in accuracy have been proposed at Fermilab by E989  and J-PARC.
First results from E989 could be available around 2017/18.


\section*{Acknowledgments}
It's a pleasure to thank the LC11 local organizing committee, particularly G. Pancheri, for running a smooth and productive meeting in a very friendly atmosphere.
I thank D. Hertzog and L. Roberts for useful discussions on the E821 and E989 experiments, 
and S. Eidelman, F.~Jegerlehner, W. Kluge, M.~Passera and L. Roberts for a careful reading 
of the manuscript. 
Support from ECT* is warmly acknowledged.

\end{document}